# Characterization of Cs-free negative ion production in the ion source SPIDER by Cavity Ring-Down Spectroscopy

**M. Barbisan,**[a,1] **R. Agnello**[a,b]**, G. Casati**[c]**, R. Pasqualotto**[a]**, E. Sartori**[a] **and G. Serianni**[a]

[a] *Consorzio RFX, corso Stati Uniti 4 – 35127 Padova, Italy*

[b] *École Polytechnique Fédérale de Lausanne (EPFL), Swiss Plasma Center (SPC), CH-1015 Lausanne, Switzerland*

[c] *Imperial College London, Exhibition Rd., South Kensington, SW7 2BX, London UK*

*E-mail*: marco.barbisan@igi.cnr.it

ABSTRACT: The Neutral Beam Injectors of the ITER experiment will be based on negative ion sources for the generation of beams composed by 1 MeV H/D particles. The prototype of these sources is currently under testing in the SPIDER experiment, part of the Neutral Beam Test Facility of Consorzio RFX, Padua, Italy. Among the targets of the experimentation in SPIDER, it is of foremost importance to maximize the beam current density produced by the accelerator. The SPIDER operating conditions can be optimized thanks to a Cavity Ring-down Spectroscopy diagnostic, which provides line-integrated measurements of negative ion density in proximity of the accelerator apertures. The specific implementation in SPIDER shows a drift in ring down time measurements, which develops in a time scale of few hours, thus possibly affecting the negative ion density estimates in plasma pulses of 1 h duration, as required by ITER. Possible causes and solutions are discussed. Regarding the source performance, this paper presents how negative ion density is influenced by the RF power used to sustain the plasma, and by the magnetic filter field present in SPIDER to limit the amount of co-extracted electrons. In this study, SPIDER was operated in hydrogen and deuterium, in Cs-free conditions.

KEYWORDS: cavity ring-down spectroscopy; negative ion source; neutral beam injector.

**Contents**

## 1. Introduction

The Heating Neutral Beam injectors (HNBs) for ITER are expected to deliver 16.7 MW beams, composed by H/D particles at 1 MeV energy. The solution adopted to produce neutrals at this energy is based on the gas neutralization of $H^-/D^-$ ions, generated by a radiofrequency (RF) inductively coupled (ICP) plasma source and accelerated by a system of grids at increasing potential [[1]-[4]]. The prototype of the ion sources for the ITER HNBs is under test in the SPIDER experiment, part of the Neutral Beam Test Facility (NBTF) at Consorzio RFX (Padua, Italy). In SPIDER, the ion source is coupled to a three-grid acceleration system, allowing to make negative ions reach up to 100keV energy. Source and accelerator are expected to deliver a beam current density of 350 $A/m^2$ (H)/285 $A/m^2$ (D) from a total extraction area of 0.2 $m^2$, and with a ratio of co-extracted electrons below 0.5 ($H^-$)/1 ($D^-$) [[1]-[4]]. With the current, technology, these performances can be only obtained relying on Cs evaporation in the source, which enhances $H^-/D^-$ production reactions on the surfaces around the extraction apertures of the acceleration system [[5],[6]]. In the first experimental campaigns of SPIDER (May 2018 - April 2021), which are object of this paper, Cs was not evaporated in the source in order to manage the main technological challenges of the source and characterize its plasma properties; negative ions were only produced by volume reactions involving vibrationally excited hydrogen/deuterium molecules [[5],[6]]. Even at reduced performance, maximizing the extracted beam current density and minimizing the co-extraction of electrons are still key targets of the tests on SPIDER. This requires to decouple the understanding of source performances in terms of negative ion production from the performance of the beam extraction and acceleration system. This is possible thanks to the Cavity Ring-Down Spectroscopy (CRDS) [7] diagnostic installed in SPIDER [[8],[9]], which provides Line-of-Sight (LoS) integrated measurements of $H^-/D^-$ density in proximity of the ion extraction apertures, as already successfully demonstrated in most other negative ion sources for fusion applications [[6],[10]-[17]]. The CRDS diagnostic in SPIDER can provide $H^-/D^-$ density measurements at 10 Hz rate, with a minimum detection threshold of few $10^{15}$ $m^{-3}$. The paper first briefly describes the structure of SPIDER and how the CRDS diagnostic was installed on it (sec. 2). Subsequently, the paper presents some technical issues about the CRDS diagnostic, which are presently under study; their resolution is important to guarantee the accuracy of $H^-/D^-$ density measurements for 1 h long plasma pulses (sec. 3). In sec. 4, the ion source performance in terms of $H^-/D^-$ density availability is studied as a function of the input RF power and of the magnetic filter field, which is applied to minimize the amount of electrons co-extracted from the source.

## 2. SPIDER and the CRDS diagnostic

A vertical cross section of the SPIDER ion source and accelerator is schematized in Figure 1a. The plasma is generated by eight cylindrical ICP plasma sources (called plasma drivers), arranged in two columns and four rows; each row is separately fed by a 1 MHz, max. 200 kW RF generator. The H/D plasma diffuses from each driver in a common expansion chamber, to uniformly reach the apertures of the three grids of the acceleration system: Plasma Grid – PG, Extraction Grid – EG and Acceleration Grid – AG. Source case and accelerator are entirely inside the SPIDER vacuum vessel (few tens of milliPascal), [[2]-[4]]. In order to minimize the co-extraction of electrons, a magnetic filter field is generated in proximity of the PG, a current (up to 5 kA) which flows in downward direction along the grid itself: 1.6 mT is produced close to the PG by 1 kA of current [[1],[5],[6],[18],[19]]. The amount of co-extracted electrons can be further reduced by modifying the potentials in the expansion chamber [[5],[6],[18]-[21]]. The ISBI (Ion Source BIas) power supply can negatively polarize the source body with respect to the PG; a second power supply, ISBP (Ion Source Bias Plate), can positively polarize the Bias Plate (BP), an electrode at 10 mm distance from the PG, open in correspondence of the PG apertures, with respect to the source body.

To detect negative ions in SPIDER [9], the CRDS diagnostic technique exploits their capability to absorb the photons of a laser pulse, because of photo-detachment reactions: $H^-/D^- +\gamma \rightarrow H/D+e$; this phenomenon is amplified by trapping the laser pulse in an optical cavity, which crosses the regions in which negative ions are present. The structure of the CRDS diagnostic is schematized in Figure 1b. In SPIDER, the laser pulses are generated outside the concrete bioshield, to protect the laser head from neutron radiation. The laser characteristics are:1064 nm wavelength, 150 mJ energy, 6 ns duration and up to 10 Hz repetition rate. The laser head is installed on an optical table inside a small room close to the bioshield; the table is fastened to the concrete walls to ensure mechanical stability. On the same optical table a visible 532 nm laser beam is co-aligned to the main one, to facilitate the process of alignment of the diagnostic. The two laser beams enter the bioshield through a dedicated aperture and reach the SPIDER vacuum vessel, where they are routed to the entrance of the optical cavity by a system of four steering mirrors. The cavity is composed by two HR (high reflectivity, >99.994 %) mirrors, installed at opposite sides of the vacuum vessel, which also act as vacuum-air interface. The mirrors must be aligned between themselves and with respect to the 10 mm diameter apertures in the source side walls; the laser pulse is reflected back and forth by the mirrors through the source volume, and partially absorbed by the $H^-/D^-$. The resulting LoS is 5 mm far from the upstream face of the PG; the mirror-mirror distance is L=4.637 m, while the LoS chord length in which negative ions are present is about d=0.612 m [9]. As schematically shown in Figure 1a, the LoS is at the top of the bottommost groups of PG apertures. The alignment of the HR mirrors is provided by vacuum tight structures, as depicted in Figure 1c. The bellows and the two dovetails systems allow for ±10 mm XY translations, while three micrometric screws indirectly push on each HR mirror and then on their O-ring to provide tilting alignment. Each alignment support is equipped with a gate valve to substitute the HR mirrors while preserving the SPIDER vacuum environment. On the opposite side of the beam injection stage in the optical cavity, a fiber collimator collects the train of laser pulses which exit the cavity, as result of the back and forth travel of the light inside the cavity. The light is then transferred via a silica, 1 mm core diameter fiber to the optical room, to an APD detector equipped with an interference filter to reject the plasma light; the detector signal is acquired at 250 MS/s.

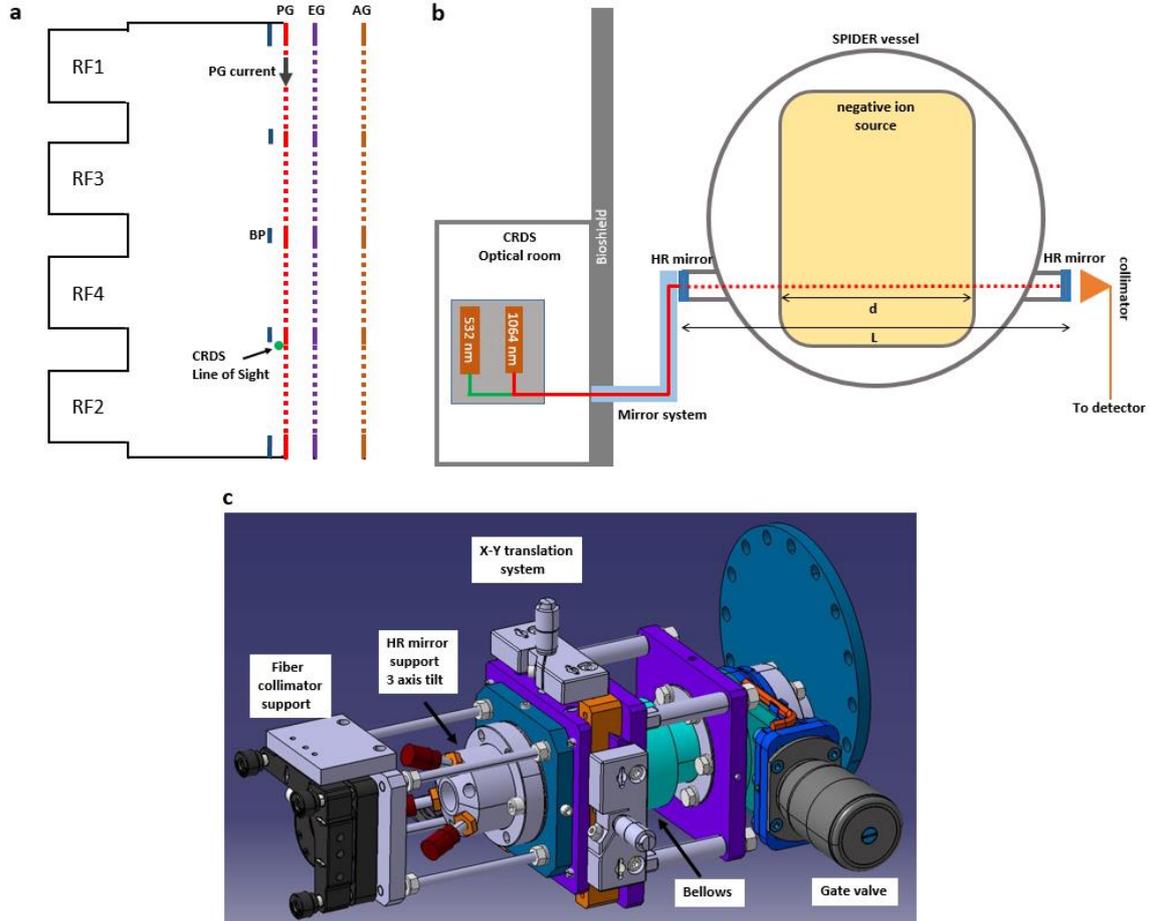

**Figure 1.** a) Scheme (not in scale) of the source and of the acceleration system of SPIDER. b) Simplified scheme of the CRDS disgnostic structure in SPIDER.c) 3D representation of the vacuum-air interface and alignment system for the high reflectivity mirror on the light reception side.

The train of light pulses which is detected after each laser pulse decays exponentially with time, with a certain ring-down decay time $\tau_0$; its value depends on the fraction $T$ of laser beam energy lost at each mirror-to-mirror travel of light into the cavity:

$$\tau_0 = \frac{L}{cT}$$

where $c$ is the speed of light The minimum value of $T$ is equal to the transmission coefficient of the cavity mirrors. The decay time can be easily measured by fitting with an exponential curve the detector ring-down signal, after a proper background subtraction. When the plasma is active, the presence of an absorbing medium (negative ions) reduces the decay time to a lower value $\tau$:

$$\tau = \frac{L}{cT'}, \text{where } T' = (T + d\sigma n_{H-})$$

and $\sigma = 3.5 \cdot 10^{-21}$ m$^2$ is the photo-detachment cross section at 1064 nm photon wavelength [22] The line averaged negative ion density can then be estimated as

$$n_{H-} = \frac{L}{\sigma c d}\left(\frac{1}{\tau} - \frac{1}{\tau_0}\right)$$

.

## 3. Ring Down time drifts

A drift of the ring-down time over RF pulses of several tens of seconds was observed since the beginning of SPIDER operations. Since this can be a relevant source of systematic error, the drift is compensated by fitting the $\tau_0$ values before and after the plasma phase over time, with a straight line; in this way $\tau_0$ can be derived for each value of $\tau$ during the plasma phase [9]. This solution proved to properly work for plasma phases and ring-down decay time drifts of the order of minutes. With the increase of SPIDER shot duration, it was however observed that the CRDS $\tau$-$\tau_0$ measurements are also subject to drifts that evolve in the time scale of hours. These drifts may represent a real issue when plasma phases will last 1 h, as required for ITER operation; even if so far the drifts have been sufficiently linear to be compensable at the end of the SPIDER shot (acquiring the final $\tau_0$ values), it would be quite difficult to have reliable feedback from the CRDS diagnostic during the plasma phase. It is then necessary to identify the source of the drifts.

Long term drifts appear as a monotonic reduction of $\tau$ and $\tau_0$ values during an experimental day; the effects of the drift disappear in between experimental sessions, during the night, so that $\tau_0$ is basically the same every morning. The fact that the ring down time decreases represents a further issue, in particular if high values of negative ion density must be measured, as expected during Cs evaporation. Figure 2a shows an example of how the ring down decay time evolves during the entire duration of an experimental day (27/05/2021 in this case). CRDS measurements are interrupted in the time interval between SPIDER shots; The spikes are the tau drops during RF on phases.. Long term drifts were firstly explained as an effect of accumulated thermal stress of the laser head during long pulses. Reducing the laser firing rate from 10 Hz to 2 Hz helped in keeping $\tau$ and $\tau_0$ high enough at the end of the day to make accurate measurements. To test this hypothesis, during 27/05/2021 the laser system was switched off on purpose (time interval 3.2h-4.2h in Figure 2a) without improvements. Thermal effects on the laser head are not the cause of the decay time drift.

In a further test, the pulse frequency of the CRDS laser was kept at 10 Hz for more than four hours; no plasma was generated in the source, the vacuum pumping system was the only plant left active. The resulting measurements of $\tau_0$ are shown in Figure 2b, as a function of time. The trend of data proves that the laser and the CRDS optical system can produce a relevant change in $\tau_0$, up to 30 %, but that this change is not monotonic. Further tests are needed to identify the cause of the described decay time drift regularly observed during each experimental day (i.e. thermal expansions effects should be investigated).

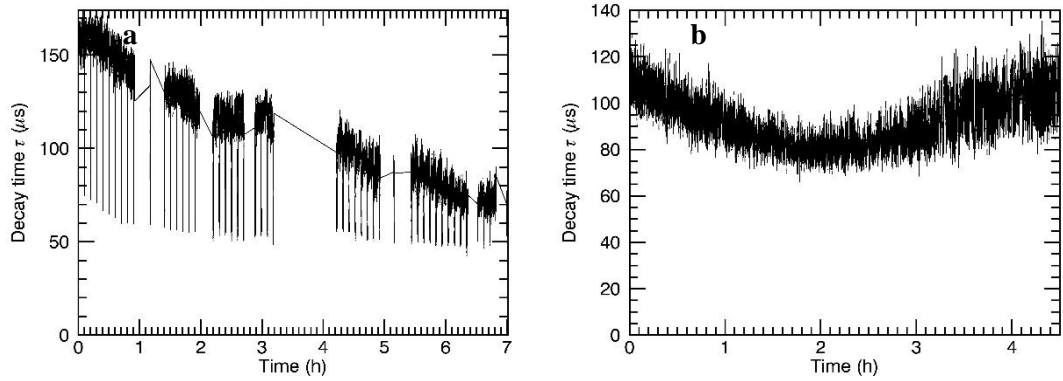

**Figure 2.** a) Ring down decay time τ as a function of time, during the 27-05-2021 experimental day. The time intervals with no data are those in between SPIDER shots. The CRDS laser firing rate was 2 Hz. b) Ring down decay time τ as a function of time, during a 4.5 h continuous 10 Hz firing of the CRDS laser. No SPIDER plant other than pumping was active.

## 4. Source experimental results

The CRDS diagnostic was used to characterize and guide the maximization of negative ion density at the PG, as a function of the source operation parameters. The volume production of negative ions increases with the plasma density, which in turn grows with the input power used to sustain the plasma [[5],[6],[10]]. Figure 3 shows the H$^-$ density increase with the total RF power, equally divided among the four RF generators. The source pressure P$_{source}$ was 0.33 Pa, the PG current I$_{PG}$ was 1.5 kA, no beam was extracted from the source. The three measurements at lowest power were taken without biasing of source components, while in the measurement at 350 kW the PG was set at 42 V above the source body voltage. In the explored total RF power range, the H$^-$ density grows with RF power; the application of a bias resulted in a slight improvement of negative ion density, with respect to the trend given by the other three cases.

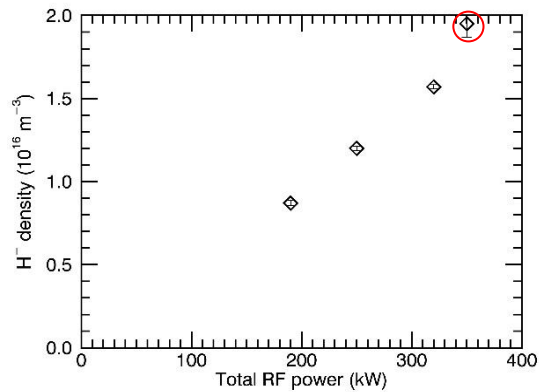

**Figure 3.** H$^-$ density as a function of total RF power; operation conditions: p$_{source}$=0.33 Pa, I$_{PG}$=1.5 kA, no biasing of source and BP (except the red circled point), no beam extraction.

The possibility to independently vary the RF power for each row of plasma drivers is useful to compensate the vertical plasma non uniformities that are expected, being the magnetic filter field horizontal [[23],[24]]. The influence of each row of drivers is however not just restricted to the portion of PG directly in front of them, nor the contribution of each pair of drivers to H- density can be considered linearly additive, despite the global behaviour shown in Figure 3. As a demonstration, in a SPIDER shot the RF generators 1, 2 and 4 were separately activated at 100 kW (RF1), 100 kW (RF2) and 120 kW (RF4) to generate a deuterium plasma in the respective pairs of drivers (RF3 was not available due to technical issues). The general operation conditions were: $p_{source}$=0.28 Pa, $I_{PG}$=1.0 kA, 45 V PG-source body bias, 43 V BP-source body bias, beam extraction with 2kV in the PG-EG gap and 20 kV in the EG-AG gap. The D- density detected at the CRDS LoS is shown as a function of time in Figure 4a; for each plasma pulse, a red shaded area indicates the time interval corresponding to stationary plasma and beam conditions. The negative ion density was predictably negligible with the farthest drivers active (RF1), while average values of $(3.40\pm0.09)\cdot 10^{15}$ m$^{-3}$ and $(4.47\pm0.09)\cdot 10^{15}$ m$^{-3}$ where measured with only RF2 and RF4 active, respectively, in the highlighted time ranges. When single pairs of drivers are active, it's then clear that only the closest drivers influence the negative ion density value at a certain vertical position at the PG. This is not true, however, if multiple RF generators are contemporarily used, as it should be in a standard NBI pulse. In a set of SPIDER plasma pulses in deuterium, the RF generators 1, 2 and 4 (100 kW/ 100 kW/ 120 kW) were simultaneously active, then RF1 was switched off; similarly to Figure 4a, the related D- density measurements are shown, as a function of time, in Figure 4b. While from the previous test a negligible change in D- density was expected, the CRDS measurement dropped by 20%, from $(1.50\pm0.01)\cdot 10^{16}$ m$^{-3}$ to $(1.23\pm0.01)\cdot 10^{16}$ m$^{-3}$ (averages over the highlighted time ranges, with stationary conditions). The operation conditions were: $p_{source}$=0.34 Pa, $I_{PG}$=1.8 kA, 44/46 V PG-source body bias, 43/44 V BP source body bias, beam extraction with 3 kV PG-EG and 24 kV EG-AG. What can be concluded is that, with multiple RF generators, the negative ion density at any point of the PG does not linearly depend on the input power of the generators and the dynamics requires further study.

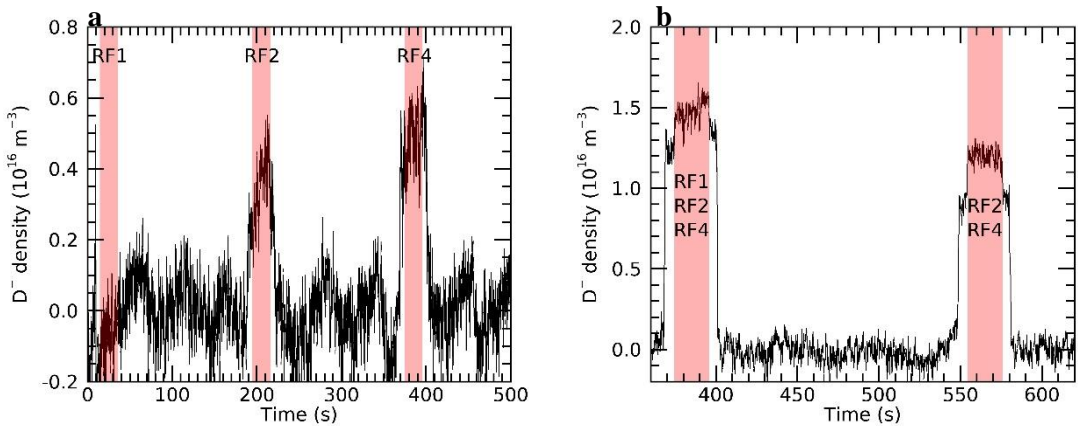

**Figure 4.** D- density as a function time, with different RF generators active. Operation conditions in plot a: $p_{source}$=0.28 Pa, $I_{PG}$=1.0 kA, 45 V PG-source body bias, 43 V BP-source body bias, beam extraction with 2kV PG-EG gap and 20 kV EG-AG gap. Operation conditions in plot b: $p_{source}$=0.34 Pa, $I_{PG}$=1.8 kA, 44/46 V PG-source body bias, 43/44 V BP source body bias, beam extraction with 3 kV PG-EG and 24 kV EG-AG. Red shaded areas indicate the time intervals with stationary plasma and beam conditions.

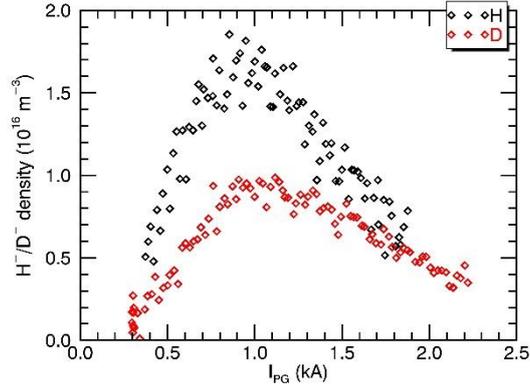

**Figure 5.** H⁻ density (black dots) and D⁻ density (red dots) as a function of the PG current $I_{PG}$. Operation conditions: $P_{RF1}$=80 kW, $P_{RF2}$=80 kW, $P_{RF3}$=0 kW, $P_{RF4}$=90 kW, $p_{source}$=0.3 Pa, no biasing of source and BP, no beam extraction.

During the Cs-free experimental campaign it was also possible to study the influence of the magnetic filter field on the density of negative ions at the PG. Figure **5** shows how H⁻ and D⁻ density vary as a function of $I_{PG}$, with no beam extraction. The source operating conditions were: $P_{RF1}$=80 kW, $P_{RF2}$=80 kW, $P_{RF3}$=0 kW, $P_{RF4}$=90 kW, $p_{source}$=0.3 Pa, no biasing of source and BP. As shown by the plot, the negative ion density is maximized at about $I_{PG}$=1 kA for both hydrogen and deuterium. In that condition, the negative ion density can be almost a factor two higher in hydrogen than in deuterium. The maximization of negative ion availability at the PG is consistent with the plasma density measurements performed by a set of movable Langmuir probes [[25],[28]] and by the fixed RF compensated probes mounted on the BP [[26],[27]]. Varying the PG current, the plasma density is peaked at about $I_{PG}$=1 kA as the negative ion density. This can be explained considering that in Cs-free conditions the negative ion production reactions (dissociative attachment of rovibrationally excited $H_2/D_2$ molecules) are driven by electrons. The decrease of the negative ion density when $I_{PG}$ is lowered from 1 kA to no current is also due to the fact that the electron temperature can rise from 2÷3 eV to 10 eV, boosting electron stripping reactions that reduce the negative ion density.

## 5. Conclusions

The negative ion source of the future ITER HNB is expected to reach challenging performances in terms of extracted beam current and duration. Maximizing the negative ion density at the apertures of the extraction system is surely one of the key targets to meet these requirements. The measurements of the CRDS diagnostic, in Cs free conditions too, gave useful information on how to manage the RF power input to the plasma, to raise both the negative ion density and its uniformity. It was also possible to find the best setting of the PG current, and then of the magnetic filter field, to optimize the density of negative ions. The drifting of $\tau_0 - \tau$ with time, which represent a concern for the accuracy of the diagnostic results long duration plasma pulses, was studied in order to identify and neutralize its causes. While the laser head can generate some still acceptable reversible variations of $\tau_0 - \tau$, thermal effects on the SPIDER vessel and the connected structure might explain the monotonic drift observed during the experimental sessions. A re-design of the CRDS mechanical structures inside the bioshield may help in reducing this phenomenon. In any case, it is necessary to provide CRDS a way to monitor the alignment of the laser beam and correlate it with $\tau_0 - \tau$, so to compensate the drift.

## Acknowledgments

This work has been carried out within the framework of the ITER-RFX Neutral Beam Testing Facility (NBTF) Agreement and has received funding from the ITER Organization. The views and opinions expressed herein do not necessarily reflect those of the ITER Organization.

This work has been carried out within the framework of the EUROfusion Consortium, funded by the European Union via the Euratom Research and Training Programme (Grant Agreement No 101052200 — EUROfusion). Views and opinions expressed are however those of the author(s) only and do not necessarily reflect those of the European Union or the European Commission. Neither the European Union nor the European Commission can be held responsible for them.

This work was supported in part by the Swiss National Science Foundation.